\documentclass[preprint2]{aastex}

\newcommand{\msun}{{\rm M_\odot}}
\newcommand{\mjup}{{\rm M_J}}

\slugcomment{To appear the Astrophysical Journal}

\shorttitle{Low-mass companions in the Solar Neighbourhood}
\shortauthors{Li, Kouwenhoven, Stamatellos \& Goodwin}

\begin{document}


\title{The long-term dynamical evolution of disc-fragmented multiple systems in the Solar Neighborhood}


\author{Yun Li}
\affil{Department of Astronomy, School of Physics, Peking University, Yiheyuan Lu 5, Haidian Qu, Beijing 100871, P.R. China\\
Center for Astronomy and Astrophysics, Department of Physics and Astronomy, Shanghai Jiao Tong University, Shanghai 200240, P.R. China}

\author{M.B.N. Kouwenhoven}
\affil{Department of Mathematical Sciences, Xi'an Jiaotong-Liverpool University, 111 Ren'ai Road, Suzhou Dushu Lake Science and Education Innovation District, Suzhou Industrial Park, Suzhou 215123, P.R.~China\\
Kavli Institute for Astronomy and Astrophysics, Peking University, Yiheyuan Lu 5, Haidian Qu, Beijing 100871, P.R. China}
\email{t.kouwenhoven@xjtlu.edu.cn}


\author{D. Stamatellos}
\affil{Jeremiah Horrocks Institute for Mathematics, Physics \& Astronomy, University of Central Lancashire, Preston, PR1\,2HE, United Kingdom}

\author{Simon P. Goodwin}
\affil{Department of Physics \& Astronomy, The University of Sheffield, Hicks Building, Hounsfield Road, Sheffield S3\,7RH, United Kingdom}



\begin{abstract}
The origin of very low-mass hydrogen-burning stars, brown dwarfs, and planetary-mass objects at the low-mass end of the initial mass function is not yet fully understood. Gravitational fragmentation of circumstellar discs provides a possible mechanism for the formation of such low-mass objects. The kinematic and binary properties of very low-mass objects formed through disc fragmentation at early times ($<10$~Myr) were discussed in \cite{li2015} (hereafter L15). In this paper we extend the analysis by following the long-term evolution of disc-fragmented systems, up to an age of 10~Gyr, covering the ages of the stellar and substellar population in the Galactic field. We find that the systems continue to decay, although the rates at which companions  escape or collide with each other are substantially lower than during the first 10~Myr, and that dynamical evolution is limited beyond 1~Gyr. By $t=10$~Gyr, about one third of the host stars is single, and more than half have only one companion left. Most of the other systems have two companions left that orbit their host star in widely separated orbits. A small fraction of companions have formed binaries that orbit the host star in a hierarchical triple configuration. The majority of such double companion systems have internal orbits that are retrograde with respect to their orbits around their host stars. Our simulations allow a comparison between the predicted outcomes of disc-fragmentation with the observed low-mass hydrogen-burning stars, brown dwarfs, and planetary-mass objects in the Solar neighbourhood. Imaging and radial velocity surveys for faint binary companions among nearby stars are necessary for verification or rejection for the formation mechanism proposed in this paper.
\end{abstract}




\section{Introduction}

Low-mass stars and brown dwarfs are among the most common objects in the Galactic field \citep[e.g.,][]{kroupa2001, chabrier2005}. The majority of the neighbours of the Sun are brown dwarfs or are of spectral type~M. Notably, the closest star to our Sun, \object{Proxima Centauri}, is an M-dwarf that orbits the \object{$\alpha$~Cen A/B} system. Our closest neighbours beyond this system are primarily of very low mass -- including \object{Barnard's star} \citep{barnard1916}, the binary brown dwarf \object{Luhman~16} \citep{luhman2013}, which may even have a third companion \citep{boffin2014}, the brown dwarf \object{WISE~0855-0714} \citep{luhman2014}, and many others, such the M-stars \object{Wolf~359} and \object{Lalande~21185}, as well as many brown dwarfs \citep[e.g.,][and numerous others]{wolf1919, ross1926, luyten1979, strauss1999, burgasser2004, burningham2010, kirkpatrick2013, troup2016}. Given their faintness, it is likely that future surveys will reveal the presence of even more nearby low-mass stars and brown dwarfs. Finally, approximately one fourth of the nearby low-mass neighbours of the Sun are known to host one or more companions \citep[e.g.,][]{burgasser2007, luhman2012, duchene2013, ward2015}, while many others are companions to higher-mass stars \citep[see, e.g.,][and references therein]{kouwenhoven2006}. 

Despite their ubiquity, the formation mechanism for low-mass objects, particularly brown dwarfs, is still poorly understood. It may be possible that brown dwarfs form from core collapse, similar to higher-mass stars \citep[see, e.g.,][]{andre2014, riaz2014, lomax2015}. However, their masses are below or close to the Jeans mass in star forming regions \citep[e.g.,][]{palau2014, degregorio2016}.  Another formation mechanism may be the gravitational fragmentation of circumstellar discs, and numerical simulations suggest that this is indeed possible \citep[e.g.,][]{stamatellos2009a, stamatellos2009b, tsukamoto2013, forgan2015, dong2016}. What fraction of circumstellar discs fragment, however, is still unknown. The decay of such disc-fragmented systems results in a population of low-mass stars, brown dwarfs and (free-floating) planetary-mass objects that contribute to shaping the low-mass end of the initial mass function \citep[see, e.g.,][]{thies2007, thies2008, thies2015, marks2015}.

Disc fragmentation results in the formation of multiple secondaries around the central star with masses ranging from the planetary to the stellar regime.   In this paper we follow the long-term (Gyr) evolution of such disc fragmented systems, using the results of \cite{li2015} (hereafter L15) as initial conditions and follow the dynamical evolution of these systems.  Throughout this paper, we follow the classification of L15 by grouping the secondaries into three categories: (i) low-mass hydrogen-burning stars (LMSs) with masses over $80~\mjup$ ($\mjup$ is the mass of Jupiter), (ii) brown dwarfs (BDs) with masses in the range $13-80~\mjup$, and (iii) planetary-mass objects (PMOs) with masses below $13~\mjup$.  We assume that all of these secondaries formed through the same mechanisms in our simulations, however each of these three categories may also form through other mechanisms \citep[see][]{whitworth2007, luhman2012}. 

L15 simulated the dynamical evolution of LMSs, BDs and PMOs formed through disc fragmentation based on the outcomes of the smoothed particle hydrodynamical (SPH) simulations of \cite{stamatellos2009a}. Their analysis covers the first 10~Myr of the dynamical evolution of these systems. They find that most systems attain a reasonably stable configuration at $t=10$~Myr, after a large number of (mostly the lowest mass) secondaries have escaped. A non-negligible fraction of secondaries have paired up into low-mass binaries, many of which escape and some of which remain in orbit around their host star. 

The simulations of L15 allow a comparison with observations of young stellar populations in or near star-forming regions and OB~associations. For a comparison with the much older field star population, however, a further analysis is necessary. In this paper we therefore carry out $N$-body simulations of disc-fragmented systems up to 10~Gyr, covering the age range of most stars in the Solar neighbourhood.  

This paper is organised as follows. We describe our methodology and initial conditions in Section~\ref{section:method}. We describe our results in Section~\ref{section:results}. Finally, we summarise and discuss our conclusions in Section~\ref{section:conclusions}.


\section{Method and initial conditions} \label{section:method}

\begin{table*}
  \begin{tabular}{lll}
    \hline
    Quantity & Set~1 & Set~2 \\
    \hline
    Mass of host star              & $M=0.7~\msun$                                  & $M=0.7~\msun$ \\
    Notional disc mass             & $M_{\rm d}\simeq 0.5~\msun$                    & $M_{\rm d}\simeq 0.2~\msun$ \\
    Number of secondaries          & $4 \le N \le 11$                               & $3 \le N \le 5$ \\
    Mass of secondaries            & $1~\mjup \leq m \leq 200~\mjup$                & $1~\mjup \leq m \leq 200~\mjup$ \\
                                   & \multicolumn{2}{l}{(at least one object with $m > 80~\mjup$)}     \\ 
    Total mass of secondaries      & $0.48~\msun \leq m_{\rm tot} \leq 0.52~\msun$  & $0.18~\msun \leq m_{\rm tot} \leq 0.22~\msun$ \\
    Semi-major axis                & $50~{\rm AU} < a < 350~{\rm AU}$ ($m<80~\mjup$)& $50~{\rm AU} < a < 250~{\rm AU}$  \\
                                   & $50~{\rm AU} < a < 150~{\rm AU}$ ($m \ge 80~\mjup$)   & \\
    Eccentricity                   & $e=0$                                          & $e=0$ \\
    Inclination                    & $0^\circ < i < 5^\circ$                        & $0^\circ < i < 5^\circ$ \\
    Longitude of the ascending node             & $0^\circ < \Omega < 360^\circ$    & $0^\circ < \Omega < 360^\circ$     \\
    Integration time               & $10$~Myr (Stage~I; L15)            & $10$~Myr (Stage~I; L15) \\
                                   & $10$~Gyr (Stage~II; this paper)                & $10$~Gyr (Stage~II; this paper) \\
    Number of realisations         & $3000$                                         & $6000$ \\

    \hline
  \end{tabular}
  \caption{Initial conditions for the two sets of simulations. The early evolution of the disc-fragmented systems up to 10~Myr is described in L15 and is referred to as Stage~I. The long-term evolution, up to 10~Gyr, is discussed in this paper and is referred to as Stage~II. The probability distributions of all parameters are described in L15. \label{table:initialtable}}
\end{table*}

We study the long-term evolution of disc-fragmented systems using $N$-body simulations, following the methodology of L15. At time $t=0$~Myr, each system in L15 initially consists of a host star of mass $0.7~\msun$ and a varying number of low-mass secondaries. L15 studied the evolution of two types of disc-fragmented systems, which are referred to as set~1 and set~2, respectively. Set~1 uses the outcomes of the simulations of \cite{stamatellos2009a} to produce the initial conditions, while set~2 corresponds to the fragmentation of lower-mass discs, and allows evaluation of the robustness of the final results on the choice of the initial conditions. The initial conditions for set~1 and set~2 are summarised in Table~\ref{table:initialtable}. 

L15 study the evolution of disc-fragmented multiple systems for the first 10~Myr. In this paper we simulate the subsequent evolution of the systems up to 10~Gyr. We refer to the simulated time span in L15 ($t<10$~Myr) as Stage~I, and the long-term evolution (10~Myr$\ <t<10$~Gyr) studied in this paper is referred to as Stage~II. Stellar populations with ages corresponding to Stage~I are typically observed in or near star-forming regions and OB associations, while Stage~II are relevant for a comparison with the field star population.

Although we continue to study the evolution of all components of the systems up to $t=10$~Gyr, it is not necessary to include the following in the $N$-body simulations: (i) host stars that are single at $t=10$~Myr, (ii) systems that have a host star with only one bound companion at $t=10$~Myr, and (iii) single or binary secondaries that have escaped before $t=10$~Myr. As the above-mentioned objects do not experience any dynamical evolution after Stage~I, they are not integrated, and are simply added to the data set after the $N$-body simulations of the other objects have finished. The final datasets thus contain 3000 systems (set~1) and 6000 systems (set~2), respectively. 

We assume that all systems evolve in isolation.  Encounters may alter the properties of multiple systems in the field, but close encounters are very rare in the low-density environment of the field ($\sim 0.1$ systems pc$^{-3}$), and distant encounters will only affect the very widest systems.

We carry out the simulations of these systems using the MERCURY6 package \citep{chambers1999}. At the end of the simulations, at $t=10$~Gyr, the energy conservation $\Delta E/E$ of all the systems are below (and usually well below) $10^{-3}$. The identification of collisions, escapers and binaries, as well as the determination of orbital elements, is carried out following the prescriptions of L15. We ignore the effects of stellar evolution, as none of the bodies in our systems has a mass high enough for stellar evolution to play a role within 10~Gyr.

We will see in Section~\ref{section:results} that at $t=10$~Gyr the vast majority of the host stars have less than two bound companions, the host stars with three companions are rare, and no host stars have more than three companions. To facilitate the discussion of our results, we refer to a host star with only a single bound companion as a {\em single companion system}. When exactly two companions orbit the host star in separate orbits, we refer to the system as a {\em double companion system}. If and only if a binary companion is in orbit around a host star, we refer to the system as a {\em binary companion system}. In the latter case, the two components of the binary are mutually gravitationally bound, and their mutual centre-of-mass orbits the host star.


\section{Results} \label{section:results}

\subsection{Dynamical decay of the disc-fragmented systems}

\begin{figure}
  \centering
  \includegraphics[width=0.45\textwidth,height=!]{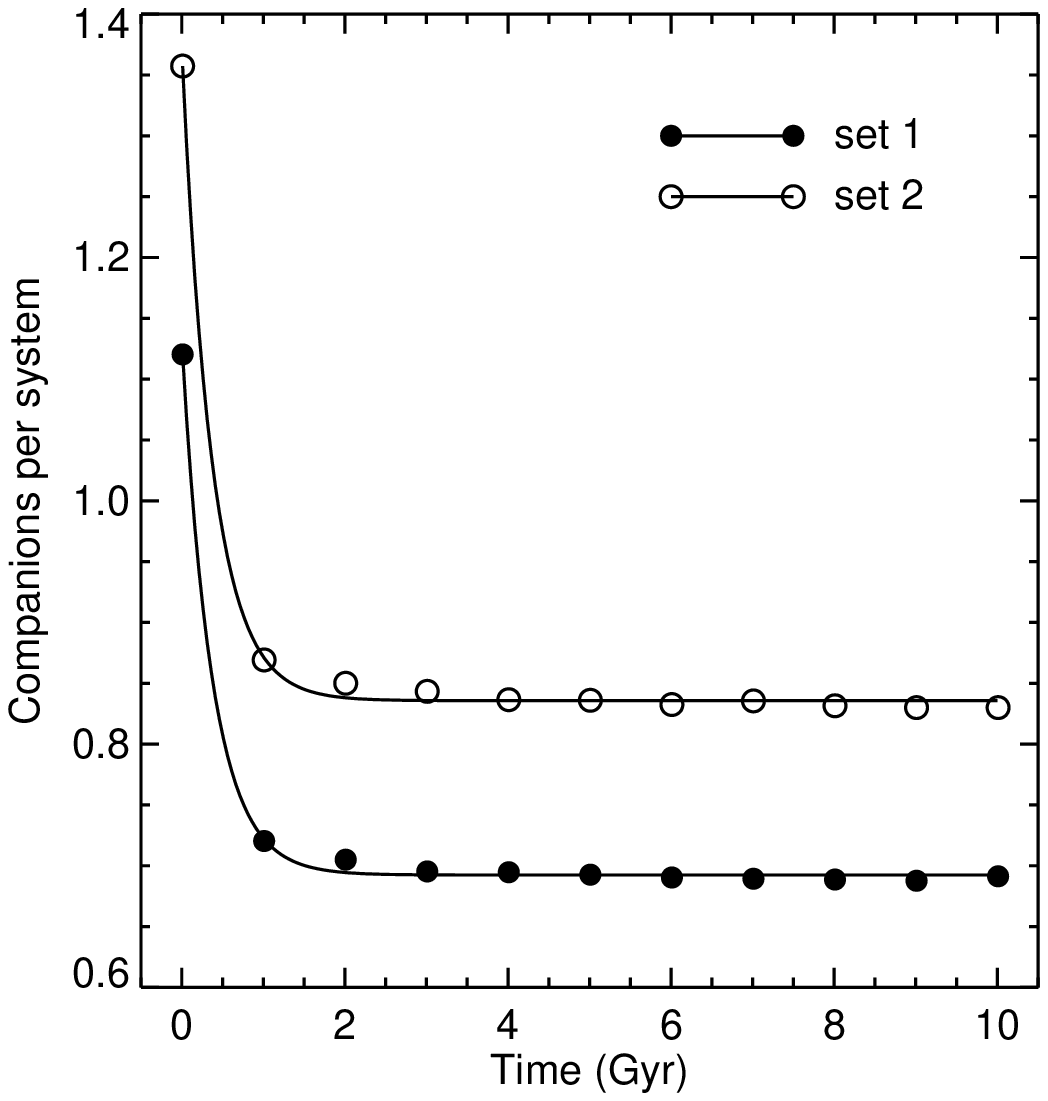}
  \caption{The number of bound secondary objects per system as a function of time, for set~1 and set~2, respectively. The solid curves represent the best-fitting exponential decay curves for each dataset.
  \label{figure:number_bound} }
\end{figure}

\begin{figure}
  \centering
  \includegraphics[width=0.45\textwidth,height=!]{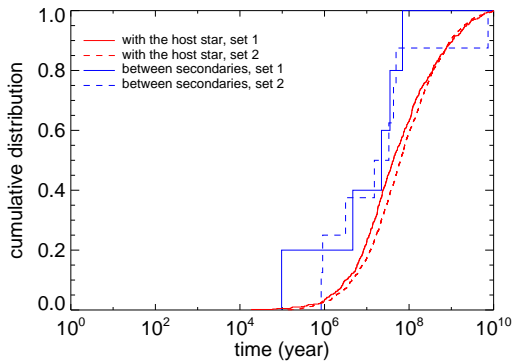}
  \caption{The cumulative distributions of the time of collisions with the host star (red) and between secondaries (blue) for set~1 (solid curves) and set~2 (dashed curves). The horizontal axis indicates the time relative to the start of Stage~II.
  \label{figure:hit_collision} }
\end{figure}

\begin{figure*}
  \centering
  \begin{tabular}{cc}
  \includegraphics[width=0.45\textwidth,height=!]{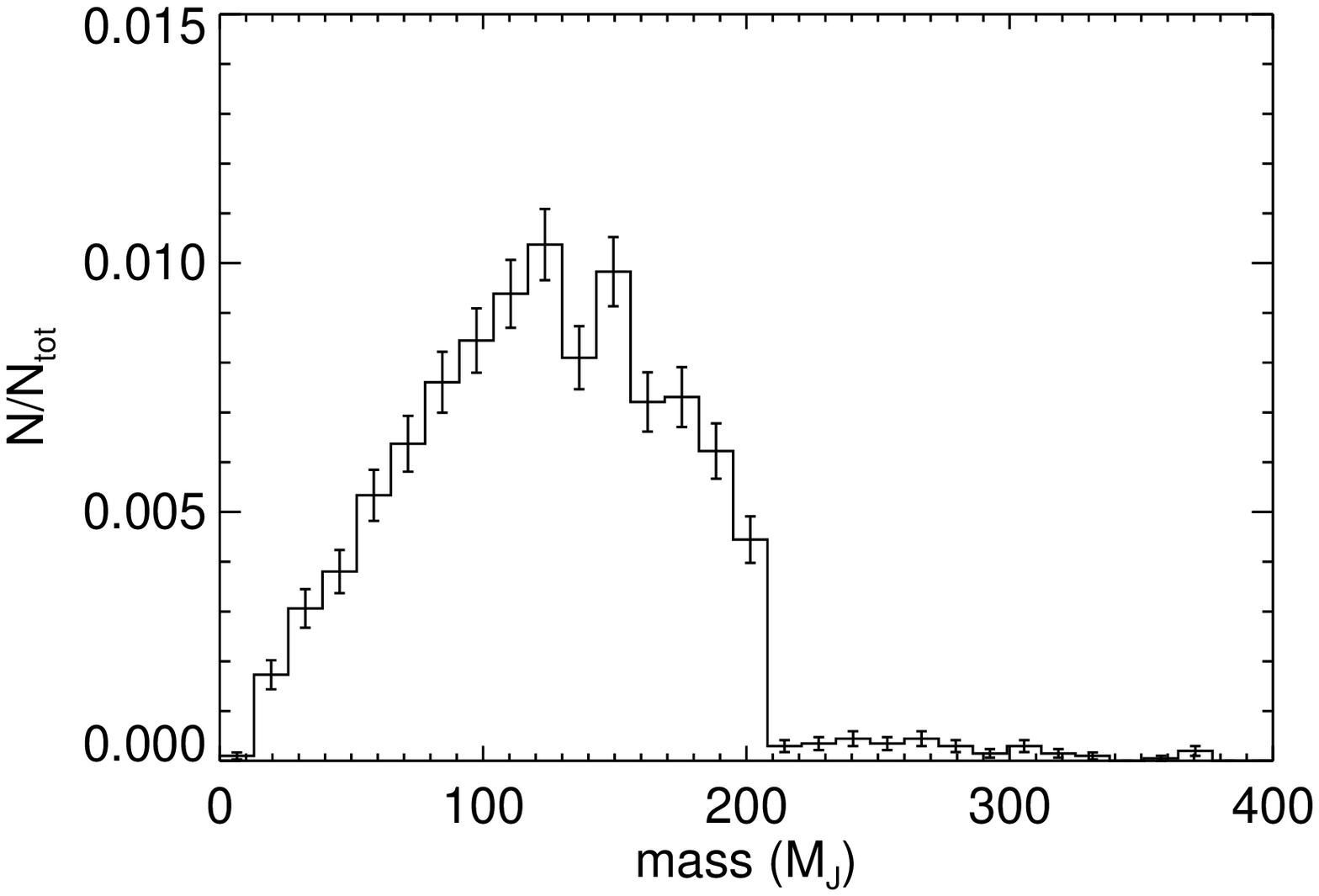} &
  \includegraphics[width=0.45\textwidth,height=!]{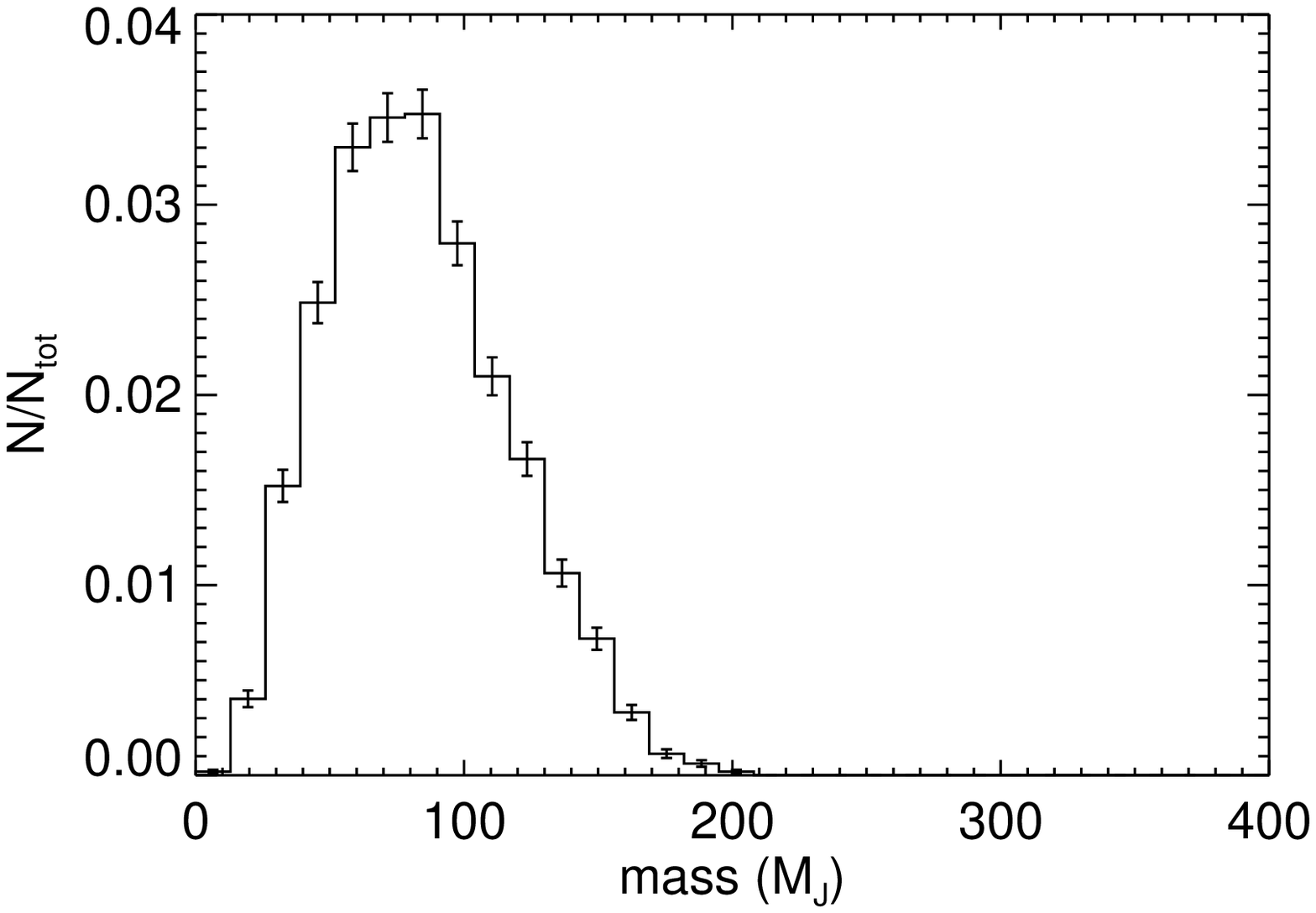} \\
  \end{tabular}
  \caption{The mass distribution of bound secondary objects at $t=10$~Gyr for set~1 ({\em left}) and set~2 ({\em right}), with Poissonian errors indicated. The distributions are normalised to the total number of secondaries at $t=10$~Gyr.
  \label{figure:m_bound} }
\end{figure*}

\begin{table*}
  \centering
  \begin{tabular}{lrr}
    \hline
    Properties at $t=10$~Gyr & Set~1  & Set~2 \\
    \hline
    Average number of companions per system   & 0.69                 & 0.83 \\
    \hline
    Single host stars                                     & $37\,\%$              & $31\,\%$ \\
    Host stars  with 1 secondary              & $57\,\%$              & $55\,\%$ \\
    Host stars  with 2 secondaries           & $5.9\,\%$              & $14\,\%$ \\
    Host stars  with 3 secondaries           & $0.13\,\%$             & $0.03\,\%$ \\
    Host stars  with $\ge 4$ secondaries     & none                  & none\\
    \hline
    Bound single companions per system                   & $0.67$            & $0.81$\\
    Bound single PMOs per system        & $<0.001$        & $<0.001$ \\
    Bound single BDs per system            & $0.14$            & $0.40$ \\
    Bound single LMSs per system          & $0.53$            & $0.41$ \\
    \hline
    Bound binary companions per system                                  & $0.011$               & $0.008$ \\
    Bound PMO-PMO binary companions per system             & none                  & none\\
    Bound PMO-BD binary companions per system                 & none                  & none\\
    Bound PMO-LMS binary companions per system              & none                   & none\\
    Bound BD-BD binary companions per system                    & $<0.001$             & $0.003$\\
    Bound BD-LMS binary companions per system                  & $0.002$               & $0.005$\\
    Bound LMS-LMS binary companions per system                & $0.008$               & $<0.001$\\
    \hline
    Fraction of PMOs bound to host star       & $0.5\,\%$       & $0.7\,\%$ \\
    Fraction of BDs bound to host star           & $3.2\,\%$        & $14.7\,\%$ \\
    Fraction of LMSs bound to host star        & $25.8\,\%$       & $67.1\,\%$ \\
    \hline
    Binary fraction among bound secondaries          & $1.6\,\% $       & $1.0\,\% $ \\
    \hline
  \end{tabular}
  \caption{Statistical properties of the systems at $t=10$~Gyr. The total number of systems modelled is 3000 for set~1 and 6000 for set~2, respectively. A comparison with table~2 in L15 shows that the systems continue to decay during Stage~II, although, given the long time span of Stage~II, the decay rate is substantially smaller than during Stage~I.\label{table:leftover} }
\end{table*}

\begin{table*}
	\centering
  \begin{tabular}{llrr}
    \hline
    \multicolumn{2}{l}{Close secondaries at $t=10$~Gyr} & set~1 & set~2 \\
    \hline
    Bound  PMOs per system      & $a \le 10$~AU    & none            & none\\
                                                         & $a \le 20$~AU    & none            & none \\
                                                         & $a \le 50$~AU    & none            & none \\
    \hline
    Bound  BDs per system        & $a \le 10$~AU    & $0.030$        & $0.0037$ \\
                                                       & $a \le 20$~AU    & $0.045$        & $0.048$ \\
                                                       & $a \le 50$~AU    & $0.051$        & $0.24$ \\
    \hline
    Bound  LMSs per system      & $a \le 10$~AU    & $0.043$        & none \\
                                                        & $a \le 20$~AU    & $0.24$        & $0.0012$ \\
                                                        & $a \le 50$~AU    & $0.37$        & $0.14$ \\
    \hline
  \end{tabular}
  \caption{The average number of close bound PMOs, BDs and LMSs per system at $t=10$~Gyr (cf. table~3 in L15). 
  \label{table:small_a} }
\end{table*}

Although the duration of Stage~I ($0-10$~Myr) is {\em much} shorter than that of Stage~II (10~Myr $-$ 10~Gyr), most of the dynamically interesting events occur during Stage~I.

For Stage~II, Figure~\ref{figure:number_bound} shows the number of (gravitationally bound) companions per system as a function of time. 
During Stage~II, the number of companions per host star decreases by 40\% in both sets of initial conditions: from 1.12 to 0.69 for set~1, and from 1.36 to 0.83 for set~2. 
The number of companions mainly decreases during the first billion years. After this time, little further evolution occurs, and the remaining systems are mostly completely stable. In most cases, the asymptotic configuration of the system corresponds to either a single host star, or a star with one companion. In a small fraction of the cases, the dynamical interaction between the companions leads to a systems with more than one companions in widely separated orbits, ensuring stability over long times.

All single host stars originate from the decay of systems in which a scattering event between two companions results in a nearly simultaneous combination of events: an ejection of one of the companions, and a collision between the other companion and the host star. An inspection of the data shows that this process is responsible for the origin of all of the 1110 single host stars in set~1 and all of the 1858 single host stars in set~2. In other words, all of the single host stars in our models are merger products. Table~\ref{table:singlestars} presents the most important dynamical processes occurring in two representative systems in which the host star ultimately ends up single.

\begin{table*}
	\centering
  \begin{tabular}{lccl}
    \hline
    Object & $m_{\rm initial}$ ($\msun$) & $m_{\rm final}$ ($\msun$) & Remarks\\
    \hline
	Star 2600 (set~1) & 0.700 & 0.913 & Merges with Comp. 1(+3) at $t=512$~kyr \\
	Comp. 1 & 0.109 & $-$   & Merges with host star at $t=512$~kyr \\
	Comp. 2 & 0.112 & 0.112 & Ejected  \\
	Comp. 3 & 0.104 & $-$   & Merges with Comp. 1 at $t=2.313$~kyr \\
	Comp. 4 & 0.083 & 0.083 & Ejected at $t=512$~kyr after scattering Comp. 1 \\
	Comp. 5 & 0.074 & 0.074 & Ejected \\
    \hline
	Star 5999 (set~2) & 0.700 & 0.798 & Merges with Comp. 3 at $t=16.75$~Myr \\
	Comp. 1  & 0.071  & 0.071 & Ejected at $t=16.75$~Myr after scattering Comp. 3\\
	Comp. 2  & 0.020  & 0.020 & Ejected  \\
	Comp. 3  & 0.098  & $-$   & Merges with host star at $t=16.75$~Myr \\	
    \hline
  \end{tabular}
  \caption{The evolution of two representative systems that fully decay into single host stars. For each of the components in the system the initial mass ($t=0$~Gyr; start of Stage~I) and the final mass ($t=10$~Gyr; end of Stage~II) are listed. The final scattering event results in the ejection of one companion and a merger between the other companion and the host star. All other ejections occur prior to this event.
  \label{table:singlestars} }
\end{table*}

Although many of the results presented in the figures and tables in this paper correspond to the final configuration at 10~Gyr, the reader should keep in mind that evolution is slow beyond 1~Gyr, and that many of these results are therefore also good approximations for  stellar population with an age spread similar to that of the Solar neighbourhood (i.e., only a small fraction of the stars is younger than 1~Gyr).

The data in Figure~\ref{figure:number_bound} can be fitted with an exponential decay function. We fit the average number of companions per system, $N(t)$, by 
\begin{equation} 
N(t)=(N_0-N_\infty)\exp(-t/\lambda)+N_\infty \quad , 
\end{equation}
where $t$ is in units of Gyr. Here, $N_0$ is the value at the start of Stage~II, and $N_\infty$ represents the value the datasets approach when time goes to infinity, which is $0.70$ and $0.84$ companions per system for sets~1 and~2, respectively. 
Note that after 10 Gyr the average number of companions per primary is less than unity (among all companions that will eventually have disappeared at 10~Gyr, the half-life survival time is $t_{1/2}=\lambda\ln 2$, which is approximately 256~Myr for both sets of initial conditions).

Physical collisions between the host star and/or companions continue to occur during Stage~II. The distribution of the collision times during Stage~II is shown in Figure~\ref{figure:hit_collision}. Almost all collisions between secondaries occur before 100~Myr, and about 90\% collisions with the host star occur within 1~Gyr. Beyond 1~Gyr, only few collisions occur, as by that time most of the remaining systems have achieved a stable configuration.  During Stage~II, each host star experiences on average 0.20 collisions with a companion in set~1, and 0.25 collisions with a companion in set~2. The number of collisions between secondaries is substantially smaller: 0.0017 per system for set~1, and 0.0013 per system for set~2. 

Compared to Stage~I, the collision rate is substantially lower during Stage~II, considering that Stage~II lasts a thousand times longer than Stage~I. The reason for the much smaller rate of collisions during Stage~II is two-fold. Firstly, there are fewer companions in the systems during Stage~II. Secondly, the systems that have survived after Stage~I are generally stable over long periods of time. Collisions with the host star occur much more frequently than collisions between secondaries. This is because the former requires a strong perturbation of one (rather than two) of the companions, while the latter requires a direct physical hit between two secondaries, which occurs less frequently.

The configurations of the systems at $t=10$~Gyr are summarised in Table~\ref{table:leftover}. By this time, almost all PMOs have escaped from their host star for both set~1 and set~2, or in rare cases, have merged with the host star or another companion. Only 3.2\% (set~1) and 14.7\% (set~2) of the BDs that formed in the fragmented circumstellar disc remain bound to the host star. As compared to the PMOs and BDs, LMSs have a substantially higher chance of remaining in orbit around the host star: 25.8\% (set~1) and 67.1\% (set~2) of the LMSs formed in the circumstellar still orbit the host star at $t=10$~Gyr. 

The relatively high retention rate in set~2, is a result of the smaller number (usually lower-mass) secondaries formed during the fragmentation process. Figure~\ref{figure:m_bound} shows the mass distribution of secondaries that remain bound up to $t=10$~Gyr. A comparison with figures~1 and~3 in L15 shows that ejection probabilities are higher for lower-mass secondaries. High-mass secondaries have a high retention rate (see also Table~\ref{table:leftover}), as ejection of such bodies almost requires a strong dynamical interaction with an even higher-mass secondary. Physical collisions between high-mass bodies result in the formation of merger products with masses beyond the range of our initial conditions ($200~\mjup$), and most of these mergers occur during Stage~I.

A comparison between the results at $t=10$~Myr (table~2 in L15) and the results at $t=10$~Gyr (Table~\ref{table:leftover} in this paper) indicates a further decay of systems with two or more companions. Although all systems initially (at $t=0$) had a large number ($3-11$) companions, roughly one third (37\% for set~1 and 31\% for set~2) of the host stars is single at $t=10$~Gyr. About half of the host stars (57\% for set~1 and 55\% for set~2) has one companion left, and will therefore remain stable for very long periods of time, although they may ultimately be disrupted following a close encounter with a neighbouring field star. 
A relatively small fraction of the systems remain in a triple: 5.9\% for set~1 and 14\% for set~2. Among these, most consist of two companions orbiting the host star, although in several occasions the two companions formed a binary that orbits the host star (see Section~\ref{section:binarycompanions}). Very few host stars are quadruples, and there are no quintuples or higher-order systems. During Stage~II the fraction of host stars with zero or one companion increases over time: all host stars with zero or one companion at the beginning of Stage~II remain as such, while other systems may decay into these. 

By the end of Stage~II approximately 1\% of the host stars is accompanied by a two companions in a binary configuration, of which the mutual centre of mass orbits the host star. The average number of single companions per system is 0.67 for set~1 and 0.81 for set~2. 
For set~1 most of the companions are LMSs, while for set~2 roughly half of the single companions are LMSs while the other half are BDs. This difference can be attributed to the smaller initial number of companions, as well as the steeper initial companion mass distribution in set~2, which has resulted in fewer dynamical ejections of BDs. The average number of PMOs is negligible in both datasets, partially because of the preferred dynamical ejections of PMOs, and partially because of the relatively small number of PMOs that have formed through disc fragmentation. 

The number of close companions per system at $t=10$~Gyr is shown in Table~\ref{table:small_a}. At $t=10$~Myr, no close ($\le50$~AU) PMOs were present, and this number remains zero throughout Stage~II. During Stage~II, for set~1 the number of close BDs decreases by a little less than 50\%, and the number of close LMSs decreases by about 25\%. As the number of close BDs decreases more strongly than the number of close LMSs, the brown dwarf desert \citep{marcy2000, grether2006, kouwenhoven2007, kraus2008, kraus2011, sahlmann2011} of set~1 becomes more prominent at the end of Stage~II than that at the end of Stage~I.


\subsection{Orbital periods and period ratios}

\begin{figure*}
  \centering
  \begin{tabular}{cc}
  \multicolumn{2}{c}{Single companion systems.}\\
  \includegraphics[width=0.45\textwidth,height=!]{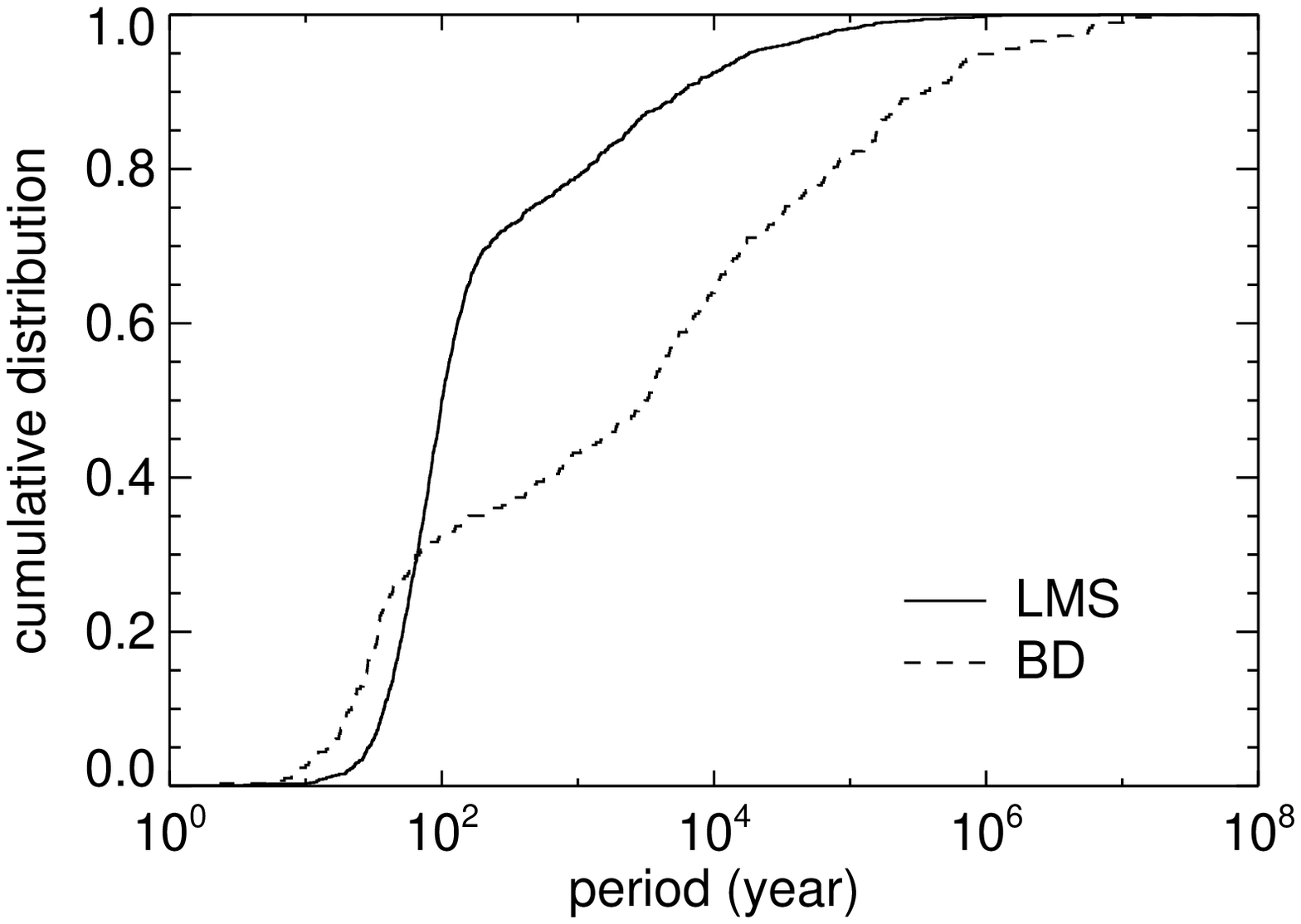} &
  \includegraphics[width=0.45\textwidth,height=!]{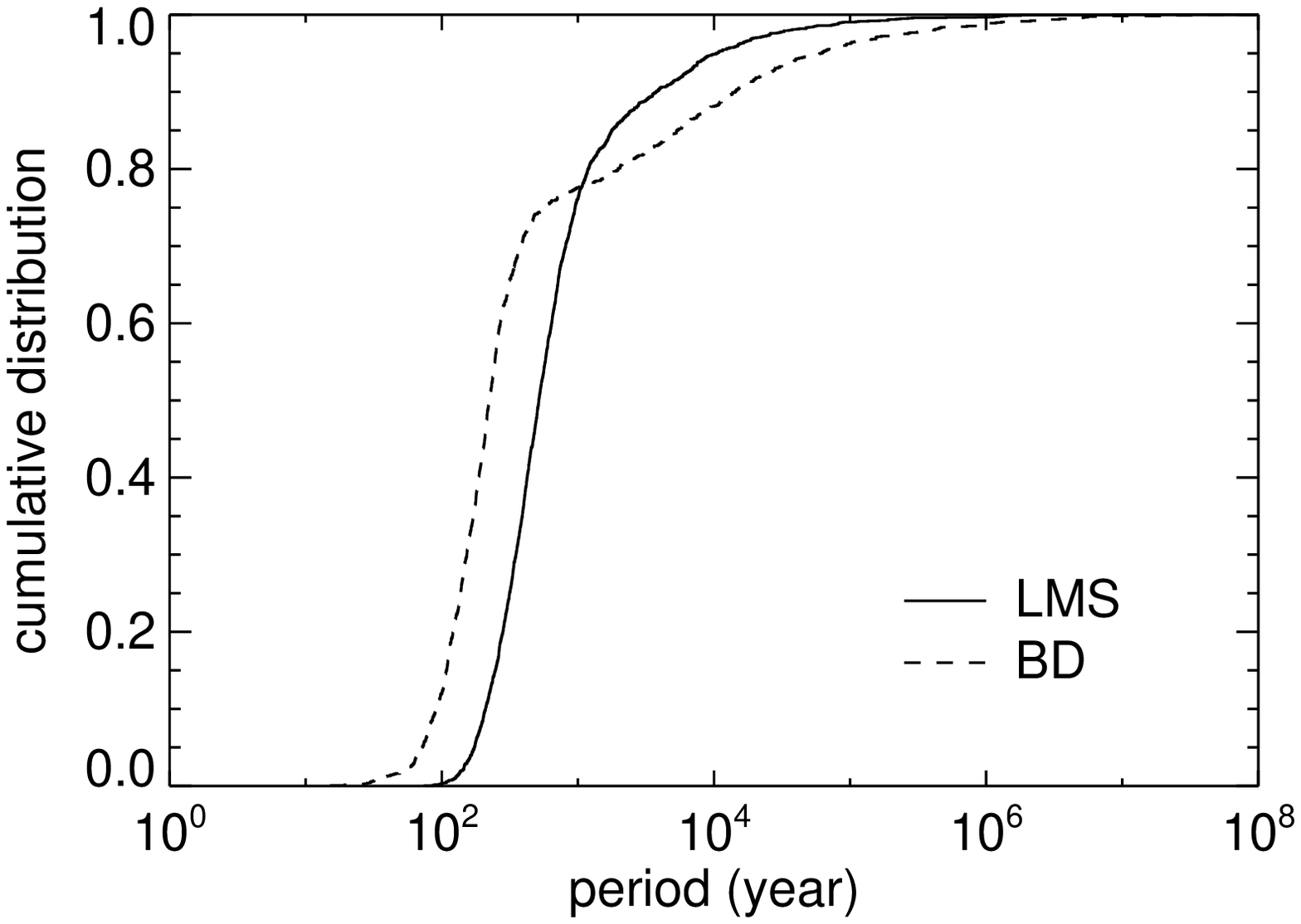} \\
\multicolumn{2}{c}{Double companion systems.}\\
  \includegraphics[width=0.45\textwidth,height=!]{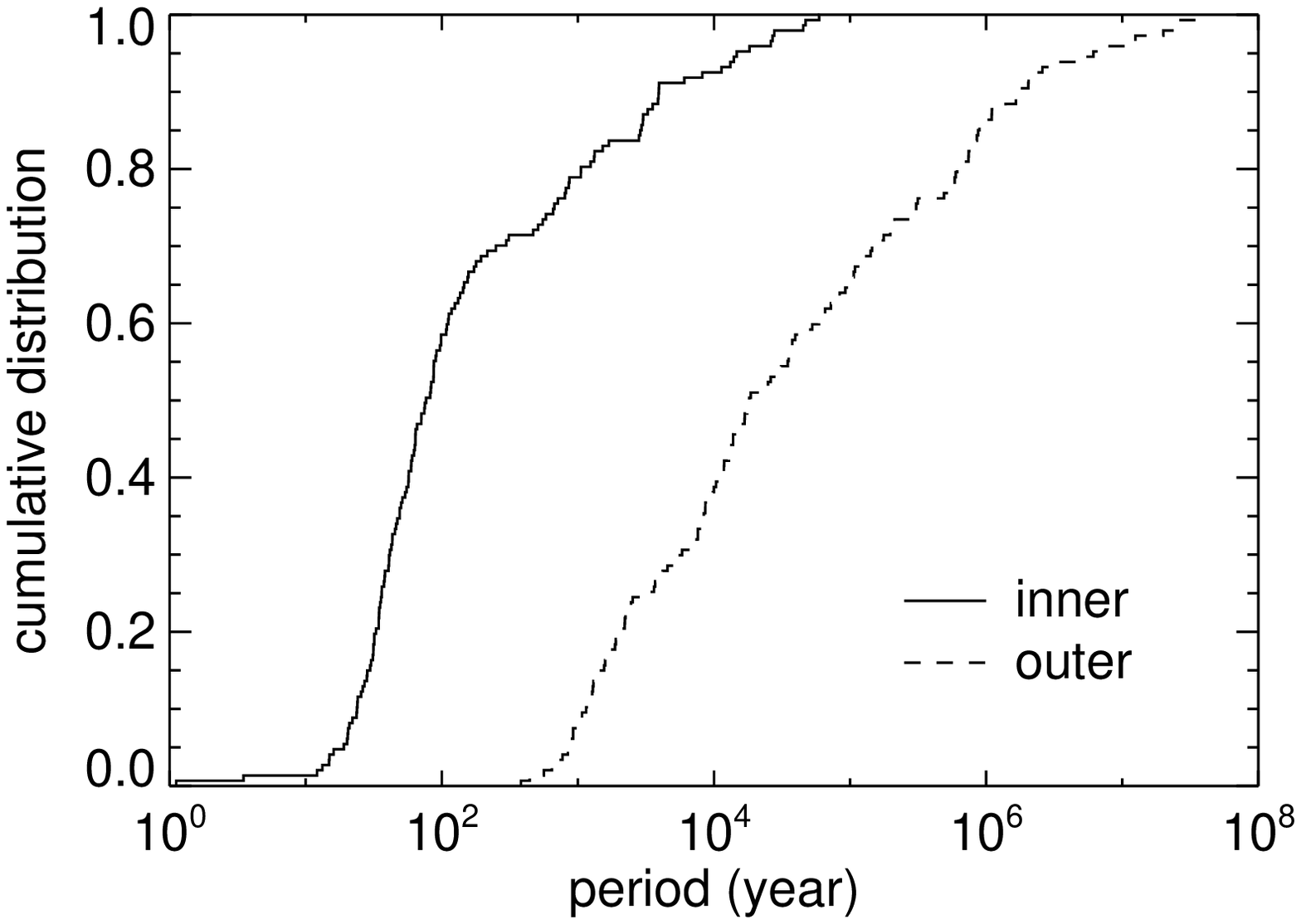} &
  \includegraphics[width=0.45\textwidth,height=!]{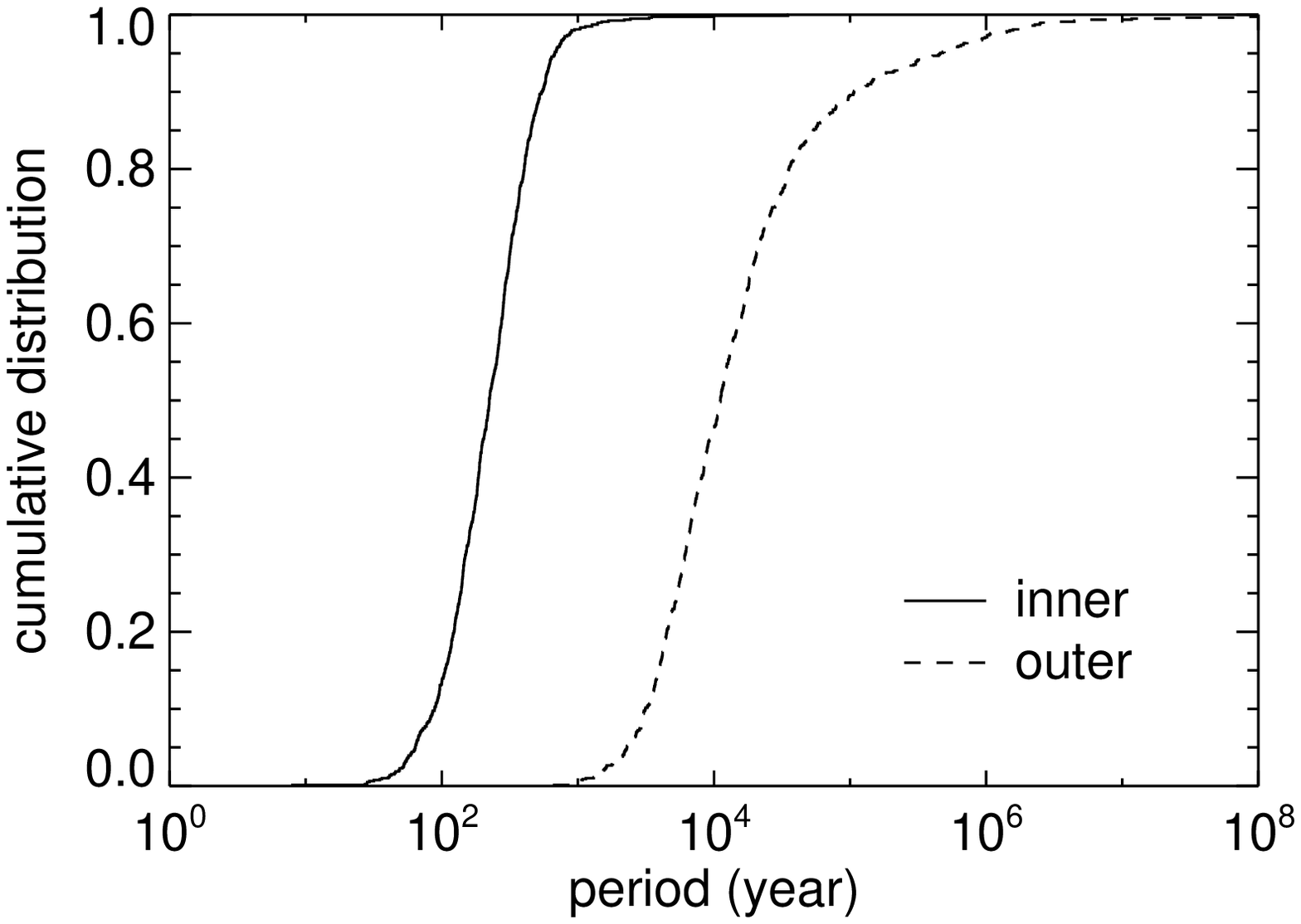} \\
\multicolumn{2}{c}{Binary companion systems.}\\
  \includegraphics[width=0.45\textwidth,height=!]{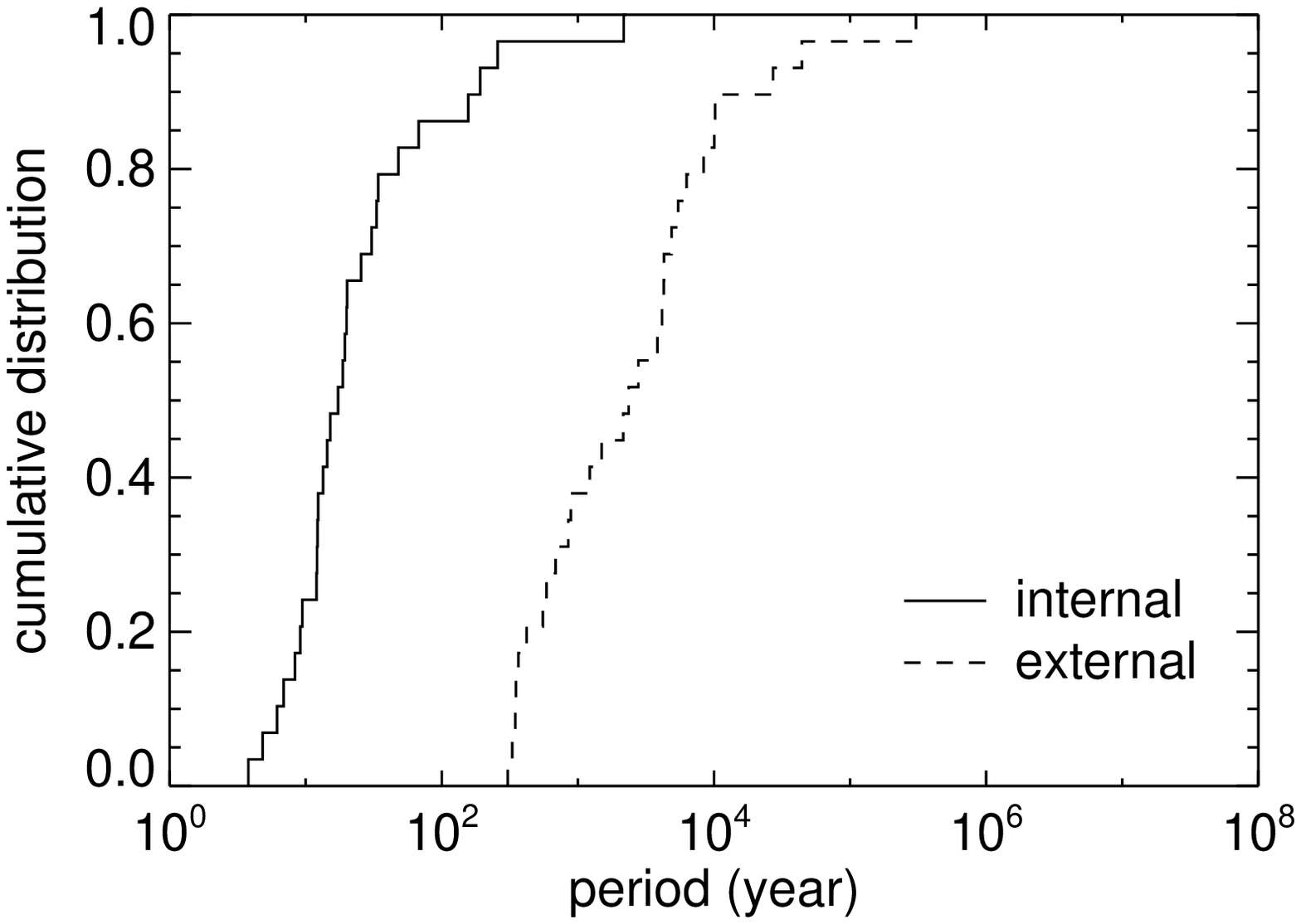} &
  \includegraphics[width=0.45\textwidth,height=!]{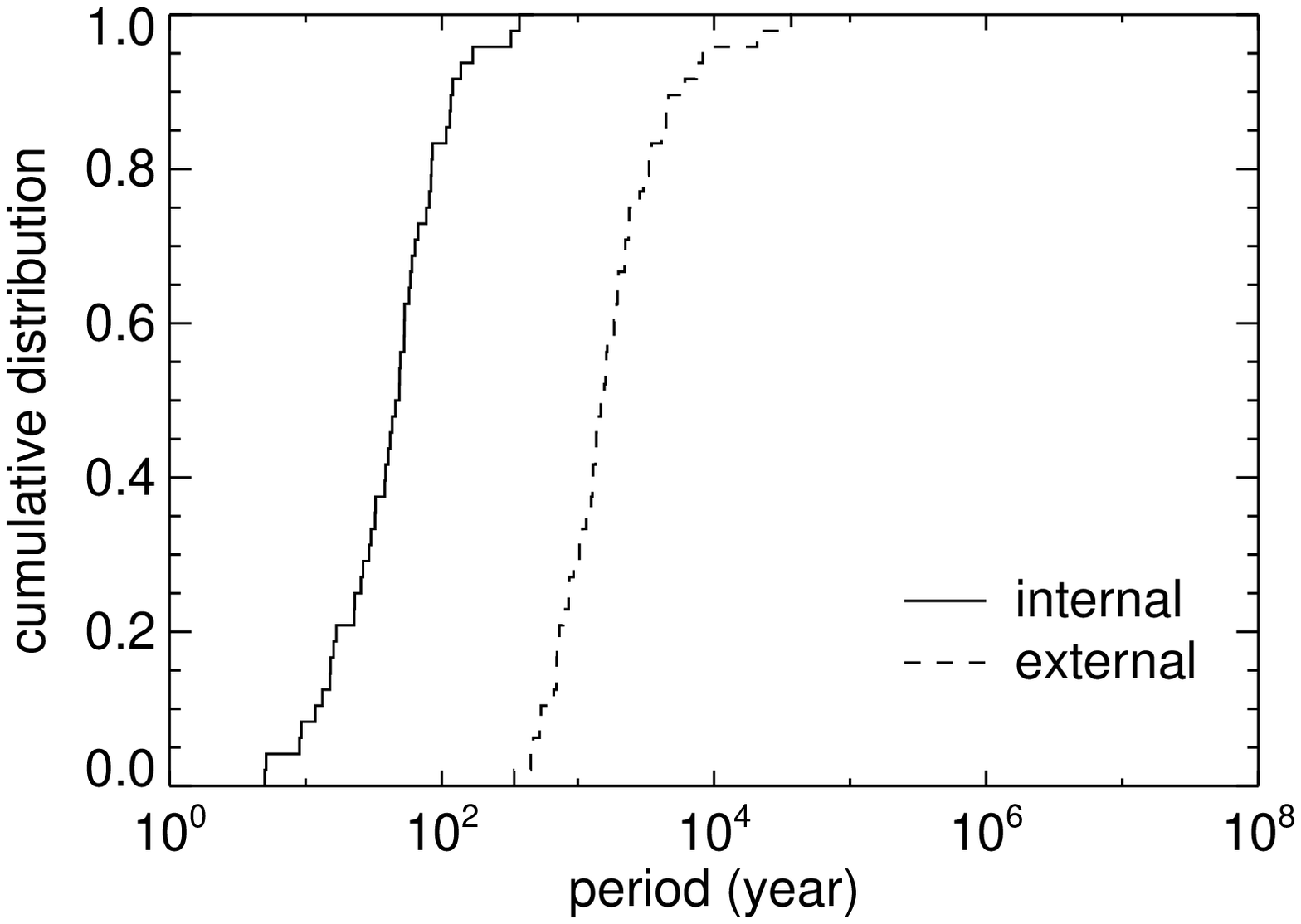} \\
  \end{tabular}
  \caption{Orbital periods of the companions at $t=10$~Gyr for set~1 ({\em left}) and set~2 ({\em right}). {\em Top:} cumulative period distributions of the LMSs (solid curves) and BDs (dashed curves) in single companion systems. {\em Middle:} cumulative period distributions of inner companions (solid curves) and outer companions (dashed curves) in double companion systems. {\em Bottom:} internal (solid curves) and external (dashed curves) cumulative period distributions in binary companion systems.
   \label{figure:period} }
\end{figure*}

\begin{figure*}
  \centering
  \begin{tabular}{cc}
  \includegraphics[width=0.45\textwidth,height=!]{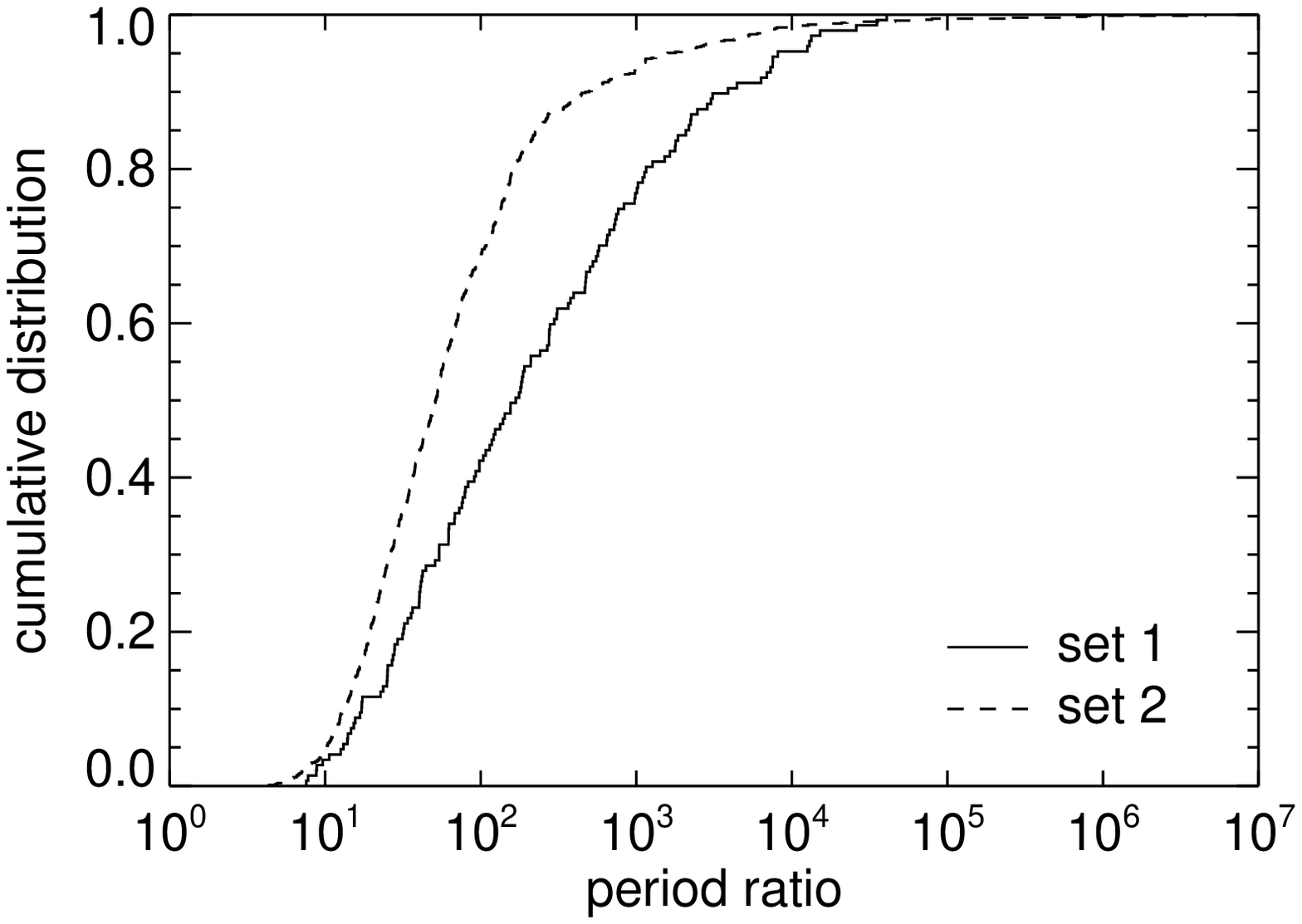} &
  \includegraphics[width=0.45\textwidth,height=!]{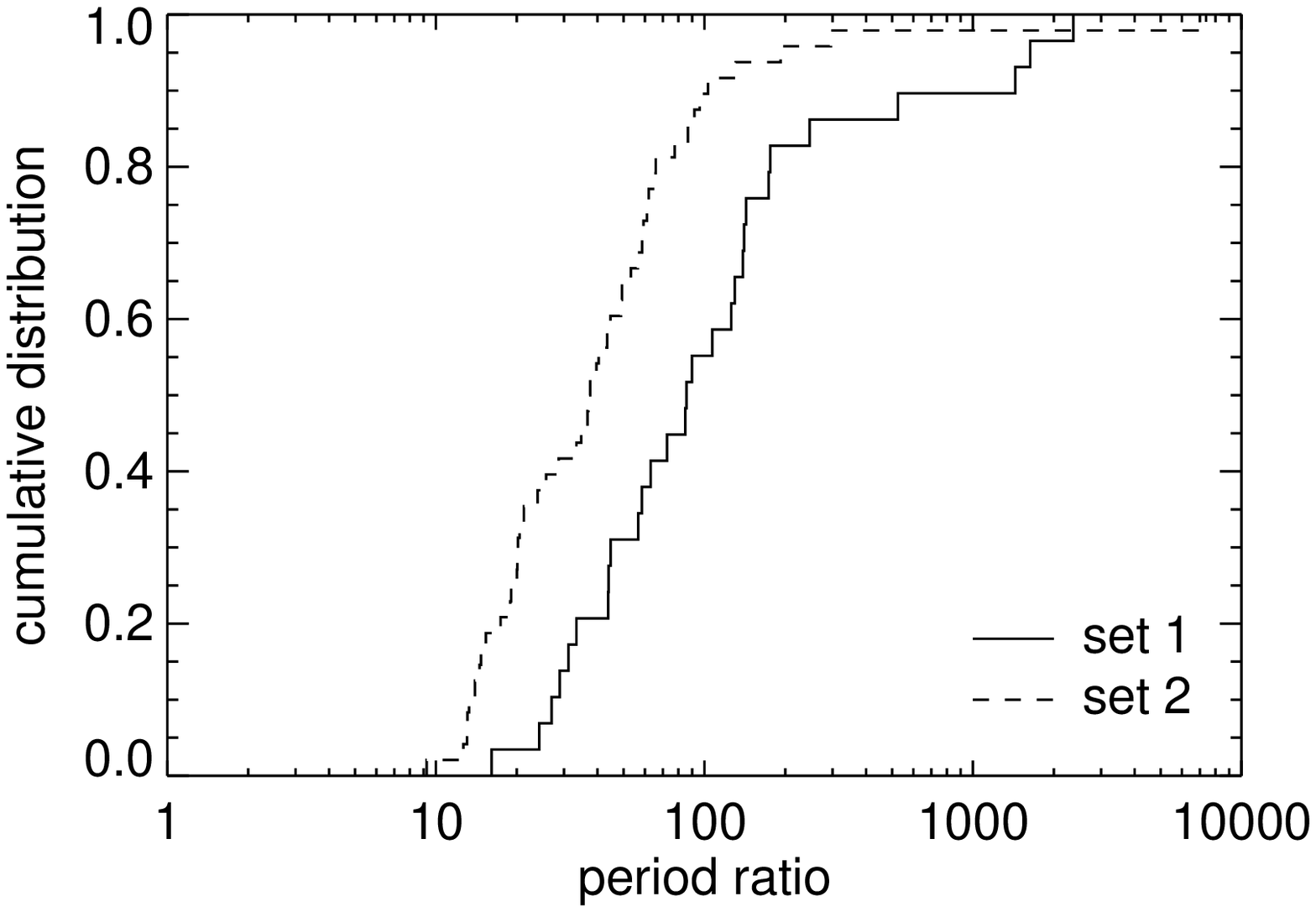} \\
  \end{tabular}
  \caption{The period ratio distribution of double companion systems ({\em left}) and binary companion systems ({\em right}) for set~1 (solid curves) and set~2 (dashed curves) at $t=10$~Gyr (cf. Figure~\ref{figure:period}).
  \label{figure:p_ratio} }
\end{figure*}

The orbital period distributions of the companions at $t=10$~Gyr are shown in Figure~\ref{figure:period}, for the three types of systems: single companion systems, double companion systems, and binary companion systems. For each system (and subsystem, when applicable) we obtain the orbital elements. This includes the {\em internal orbital elements} of the orbit of two secondaries orbiting a mutual centre of mass, and the {\em external orbital elements} of this mutual centre of mass around the host star.

In the single companion systems, most LMSs tend to have shorter orbital periods as compared to the BDs in set~1, while the periods of the LMSs and BDs are very similar in set~2, which is a result of the different initial semi-major axis distributions. For set~1, the largest initial semi-major axis for LMSs is 150~AU, but the value for BDs is 350~AU. For set~2, on the other hand, the LMSs and BDs have a nearly identical initial semi-major axis range. 

In the double companion systems, about 80\% of the inner companions have periods within the range $P_{\rm in}=10-1000$ years, and about 80\% of the outer companions have orbital periods between $P_{\rm out}=1000$~years and $P_{\rm out}=1$~Myr, for set~1. For set~2, about 90\% of the inner companions have periods in the range $P_{\rm in}=100-1000$ years, and about 90\% of the outer companions have periods between $P_{\rm out}=1000$~years and $P_{\rm out}=0.1$~Myr. 

In the binary companion systems, about 85\% of the internal periods are between $P_{\rm int}=1$~year and $P_{\rm int}=100$~years, and about 85\% of the external periods are in the range $100-10\,000$~years, for set~1. About 75\% of the internal periods are between $P_{\rm int}=10$ and $P_{\rm int}=100$ years, and about 95\% of the external periods are between $P_{\rm ext}=100$~years and $P_{\rm ext}=10\,000$~years, for set~2.

The long-term stability of hierarchical systems consisting of three or more components is to first order determined by how well the different orbits in the system are gravitationally separated from each other.
Figure~\ref{figure:p_ratio} shows the orbital period ratio distribution of double companion systems and binary companion systems. About 95\% of the period ratios of double companion systems are between $P_{\rm out}/P_{\rm in}=10$ and $P_{\rm out}/P_{\rm in}=10\,000$, which means that the inner and outer orbits are usually very well separated. About 95\% of the period ratios of binary companion systems are in the range $P_{\rm ext}/P_{\rm int}=10-1000$. These period ratios are generally large enough to ensure stability of these systems over billions of years.


\subsection{Binarity among companions in multiple systems} \label{section:binarycompanions}

\begin{figure}
  \centering
  \begin{tabular}{cc}
  \includegraphics[width=0.45\textwidth,height=!]{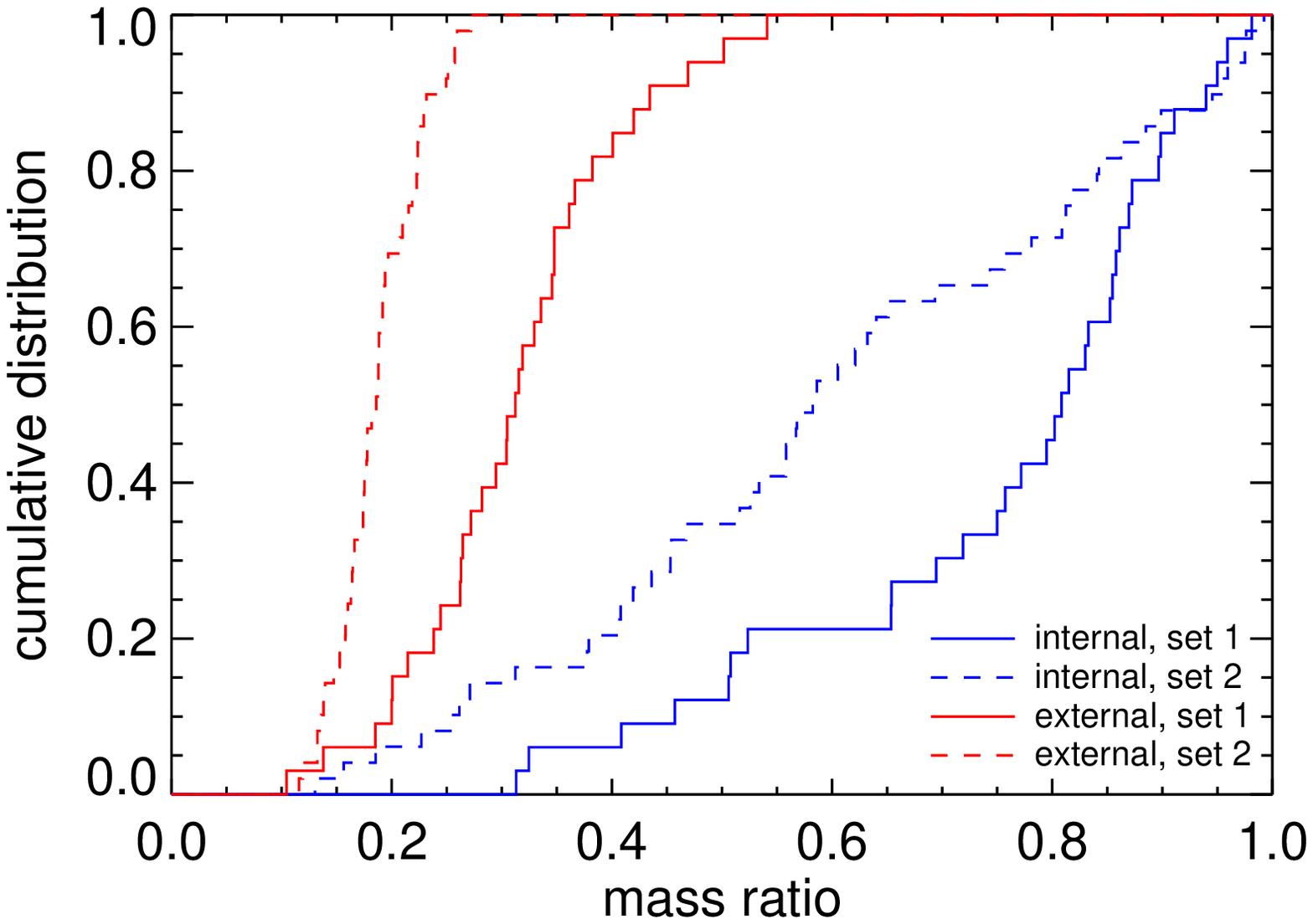} \\
  \includegraphics[width=0.45\textwidth,height=!]{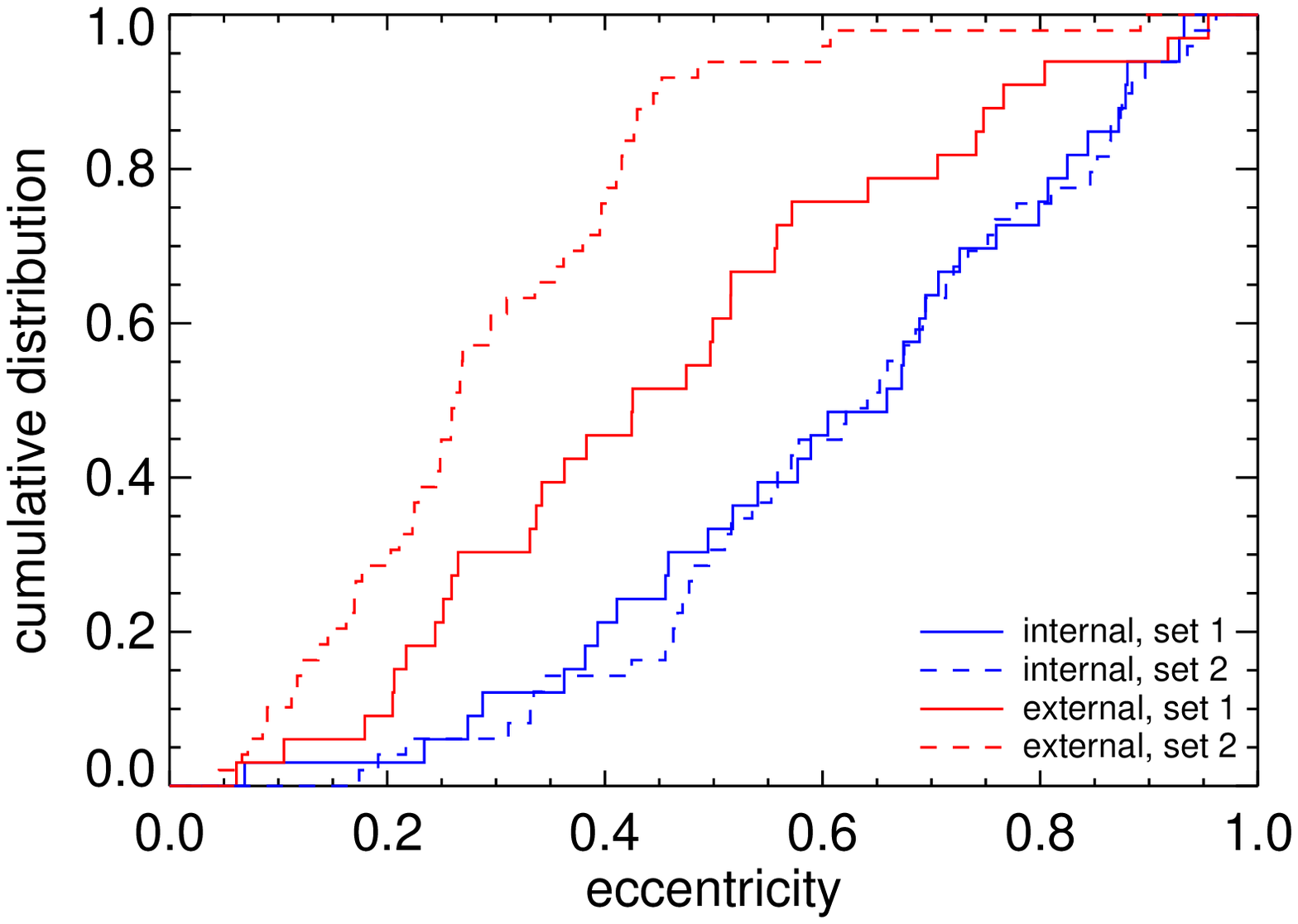} \\
  \includegraphics[width=0.45\textwidth,height=!]{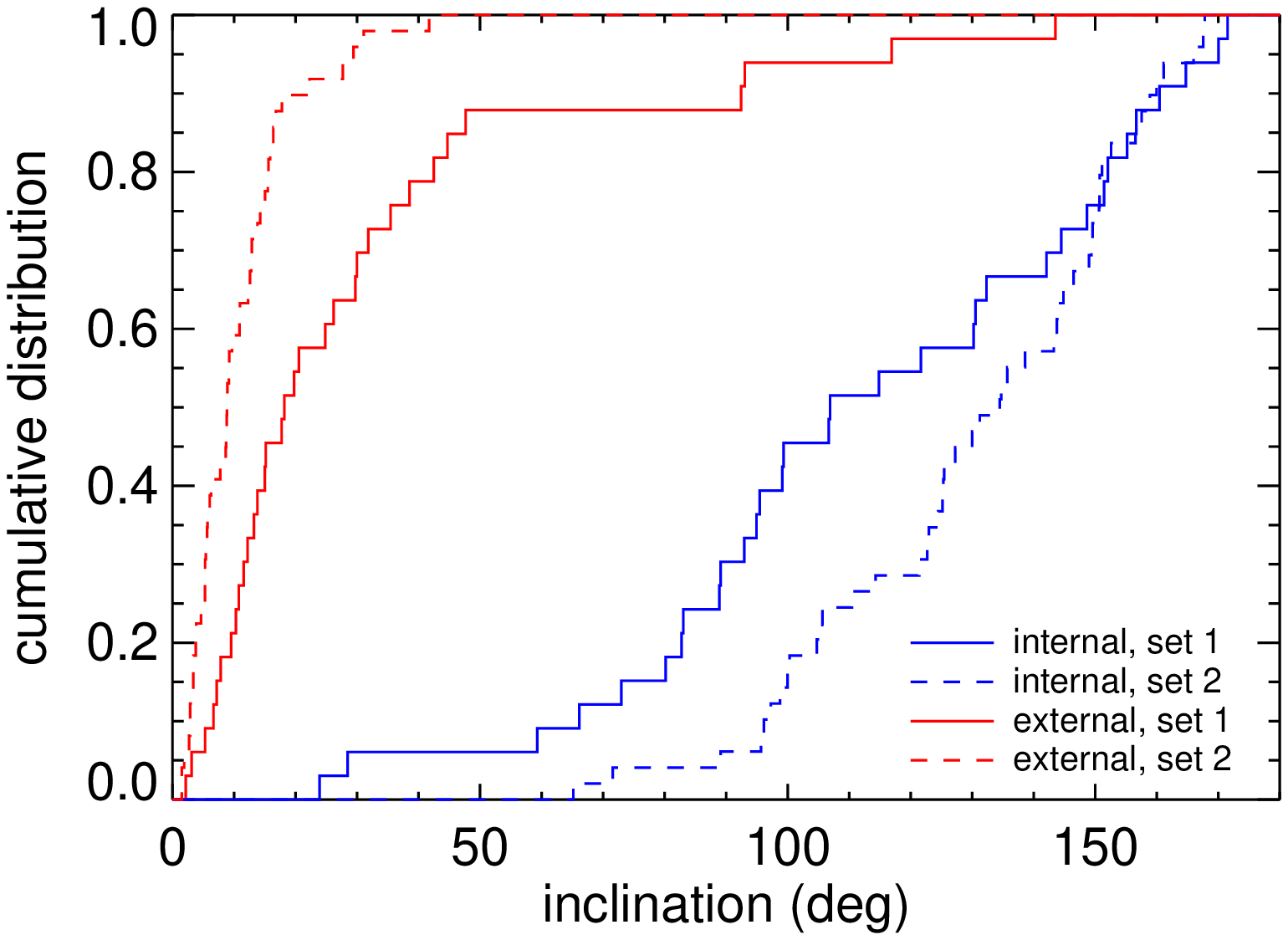} \\
  \end{tabular}
  \caption{Cumulative orbital element distributions of the systems containing a bound binary companion: the mass ratios ({\em top}), binary eccentricities ({\em middle}) and binary inclinations ({\em bottom}) for set~1 (solid curves) and set~2 (dashed curves) at $t=10$~Gyr. Blue and red curves represent internal and external results, respectively.
  \label{figure:binary} }
\end{figure}

Although binary companions are not very common at $t=10$~Gyr, they are dynamically very interesting.  The internal and external distributions of the mass ratios, eccentricities and inclinations of these binary companions are shown in Figure~\ref{figure:binary}, both for set~1 (solid curves) and set~2 (dashed curves).

For the systems with host star mass $M$ and binary companion components masses $m_1$ and $m_2$ (with $M > m_1 \ge m_2$) we define the internal mass ratio as $q_{\rm int} \equiv m_2/m_1$ and the external mass ratio as $q_{\rm ext}\equiv (m_1+m_2)/M$. As the secondaries are generally of much lower mass than the host stars, the internal mass ratio distributions are peaked at higher values than external mass ratio distributions for both set~1 and set~2. 
The median internal mass ratio is $q_{\rm int}\approx 0.8$ for set~1 and $q_{\rm int}\approx 0.6$ for set~2, respectively. The initial conditions and subsequent dynamical evolution are responsible for the higher values of both the internal and external mass ratio distributions in set~1 with respect to those in set~2.
From an observational point of view, these relatively high mass ratios mean that these binary companions should be relatively easily observable. The external mass ratios are typically $q_{\rm ext}=0.15-0.40$ and the total mass of the binary companions is therefore typically $m_1+m_2 = (0.1-0.3)~\msun$. 

The internal and external eccentricity distributions of the bound binary companions cover the whole range of eccentricities, with the exception of very small ($e\la 0.05$) and very large ($e\ga 0.95$) values. The internal eccentricity distributions $f(e_{\rm int})$ of both set~1 and set~2 are similar to the thermal eccentricity distribution $f(e_{\rm int}) = 2e_{\rm int}$ \citep{heggie1975}. The external orbits are less eccentric, which is a result of the relatively large amount of orbital angular momentum that has to be conserved when two companions pair into a binary. Other secondaries interacting with the two secondaries in a binary companion during or after its formation may remove angular momentum from the binary companion. Hence, set~1, which has more companions than set~2 at the start of Stage~I, displays an external eccentricity distribution with, on average, higher eccentricities. 

The bottom panel in Figure~\ref{figure:binary} shows the internal inclination distributions $f(i_{\rm int})$ and external inclination distributions $f(i_{\rm ext})$ for the binary companions in set~1 and set~2. All inclinations are measured with respect to the plane of the disc in which the secondaries were formed. 
For a hypothetical ensemble of completely randomly oriented orbits in space, the inclination distribution is $f(i)=\frac{1}{2}(1-\cos i)$. In such configurations, half of the orbits have inclinations larger than $90^\circ$, i.e., they are retrograde. A brief inspection of Figure~\ref{figure:binary} shows that both the internal and external inclination distributions are far from random. Most of the binary companions orbit their host star close to the plane of the disc in which they formed: approximately half of the external inclinations of the binary companions are below $i_{\rm ext}=20^\circ$ in set~1 and half are below $i_{\rm ext}=10^\circ$ in set~2. Again, the differences between the two datasets are a results of the larger initial number of secondaries in set~1 that resulted in more dynamical interactions, and therefore typically higher inclinations. Most of the binary companions have internal inclinations larger than $i_{\rm int}=90^\circ$, which means that these binaries have retrograde orbits with respect to their orbit around the host star. This was also observed at the end of Stage~I (see L15), and is mainly due to the fact that these retrograde binary companions form more easily because of angular momentum conservation. In addition, these systems are more stable than prograde binary companions, as can be seen by comparing the bottom panel in Figure~\ref{figure:binary} with figure~15 in L15. The large spread in both the internal and external inclinations of the binary companions during Stage~II suggests that for a subset of the systems the Kozai mechanism \citep{kozai1962} is responsible for large variations in both eccentricity and inclination, partially accounting for the further disruption of the disc-fragmented systems at times beyond 1~Gyr. Therefore, non-alignment of the internal orbits of hierarchical triple  system does not necessary provide evidence against formation in a disk. Thus, triple systems with orbits on the same plane, like e.g. \object{Kepler-444} \citep{dupuy2016} may not be that common, even if a large fraction of triple systems form by disk fragmentation.


\section{Conclusions and Discussion} \label{section:conclusions}

We have studied the long-term evolution of disc-fragmented systems, in order to study the orbital configurations of LMSs, BDs and PMOs orbiting solar-type stars.  This extends the simulations of L15 to cover 10~Gyr of dynamical evolution, which allows us to compare the products of the disc-fragmentation process with field population in the Solar neighbourhood. We refer to the time span studied by L15 as Stage~I ($0-10$~Myr) and the time span in this paper as Stage~II (10~Myr$-10$~Gyr). Stage~I roughly represents the period of time that the systems spend in or near their natal environment (star-forming regions and OB associations), while Stage~II allows us to compare our results with the stellar population in the Galactic field. Our main conclusions can be summarised as follows:

\begin{enumerate}
\item Systems continue to decay beyond 10~Myr due to the decay of higher-order systems, and also collisions between members of the system.  Almost all of this dynamical evolution occurs in the first Gyr leaving a very stable population after this time.

\item For approximately one third of the primaries  (37\% in set~1 and 31\% in set~2), the host star ends up as a single star, despite the large ($3-11$) number of secondaries present during the phase of disc fragmentation. More than half of the host stars have one low-mass companion at $t=10$~Gyr, while 6\% (set~1) to 14\% (set~2) of the host stars have two companions. Only a small fraction ($\la 0.1\%$) of the host stars have three companions left, while no host star is able to retain four or more of its companions.

\item The number of physical collisions with the host star is non-negligible during Stage~II. On average, each host star experiences 0.20 collisions in set~1, while the value is 0.25 for set~2. On the other hand, physical collisions between secondaries are very rare (less than 0.002 collisions per system).

\item For all primaries that ultimately end up as a single star (37\% in set~1 and 31\% in set~2), the final dynamical process to occur is a scattering event involving two companions, which results in the dynamical ejection of one of the companions, and a merger between the host star and the other of the companions. All single host stars in our models are merger products.

\item At 10~Gyr, most of the remaining single companions orbit their host star in wide orbits with periods between $100$ years and $1$~Myr (Figure~\ref{figure:period}). Very low-mass secondaries are potentially observable with imaging and radial velocity surveys. PMOs with separations less than 50~AU are absent, although higher-mass BDs and LMSs occur more frequent (ranging between, on average, 0.05 and 0.37 per system). The double companion systems contain two companions which have widely separated orbits, with period ratios mostly ranging between $P_{\rm out}/P_{\rm in}=10$ and $P_{\rm out}/P_{\rm in}=10^4$.

\item Binary companion systems orbiting the host star at $t=10$~Gyr have external-to-internal period ratios mostly ranging between $P_{\rm ext}/P_{\rm int}=10$ and $1000$. The binary companion masses are usually comparable, and the mass ratio of the binary system with respect to the host star is typically between $q_{\rm ext}=0.15$ and 0.40. The binary companions show a more or less thermal internal eccentricity distribution ($e_{\rm int}$), while their external eccentricities ($e_{\rm ext}$) tend to be more circular (although high eccentricities also exist among these external orbits). About half of the external inclinations of the binary companions are below $i_{\rm ext}=20^\circ$ in set~1 and $i_{\rm ext}=10^\circ$ in set~2, while a large majority of the binary companions have internal inclinations ($i_{\rm int}$) beyond $90^\circ$, i.e., they have retrograde orbits.

\end{enumerate}

Most nearby stars are low-mass stars or brown dwarfs that formed billions of years ago. Our study allows us to speculate somewhat on how these low-mass neighbours may have formed and evolved over time. 

Disc fragmentation provides a possible (but certainly not the only) solution for the origin of many of these. The solar system itself is clearly a result of planet formation through core-accretion. Also our closest neighbour, the \object{$\alpha$ Centauri} triple system, has likely formed differently, with its very low-mass companion \object{Proxima Centauri} perhaps being either a result of a capture event \citep[e.g.,][]{kouwenhoven2010, moeckel2011, parker2014} or a result of a triple decay event \citep[e.g.,][]{reipurth2012}. The other known objects with a distance smaller than that of \object{Sirius} are the very low mass single objects \object{Barnard's Star} \citep{barnard1916}, \object{WISE~0855-0714} \citep{luhman2014}, \object{Wolf~359} and \object{Lalande~21185}, and the very low-mass binary system \object{Luhman~16} \citep{luhman2013}. The nearby population beyond \object{Sirius} is also dominated by very low-mass objects. Although disk fragmentation may not be the dominant scenario responsible for the origin of this low-mass population, it does explain many of its properties, including the origin of the very low-mass binaries and multiples in the proximity of the Sun, such as \object{Luhman~16}, \object{Luyten~726-8}, \object{EZ~Aquarii}, \object{Struve~2398}, \object{Gloobridge~34}, and \object{Epsilon~Indi}.

Our study provides insight into a possible formation scenario for LMS, BD and PMO companions to solar-type stars in the more distant Galactic field, as well as their free-floating siblings, both in the form of singles and binaries. Many stars in the Galactic field are part of a binary or of a hierarchical multiple system \citep[e.g.,][]{tokovinin2002, tokovinin1997, tokovinin2008, tokovinin2014}. As dynamical capture is rare, these systems are almost certainly formed as such. Our theory provides an excellent explanation for the origin of several hierarchical multiples, for example, the hierarchical triple systems \object{HIP68532} and \object{HIP69113}, which both consist of a main sequence star, with a double companion made up of two low-mass stars \citep[see figure 9 in][]{kouwenhoven2005}. The scarcity of close-in brown dwarf companions predicted by our model is also reflected in observations \citep[e.g.,][]{kouwenhoven2007}. 

A deeper analysis and more detailed comparison with observations is necessary, as with our simulations we have only covered a subset of all possible initial conditions that may lead to disc fragmentation. In addition, we have not taken into account the effect of encounters with neighbouring stars and brown dwarfs, an affect that may be particularly important during first few million years, when a disc-fragmented system is still in its dense natal environment where the circumstellar disc is initially exposed to the violent interstellar medium \citep[e.g.,][]{bik2010}, and subsequently participates in rapid exchange of energy with its neighbours \citep[see, e.g.,][and numerous others]{allison2009}. Stellar encounters can disrupt existing stellar and planetary systems \citep[e.g.,][]{zheng2015, wang2015a, wang2015b, wang2016}, although in the Galactic field this mostly affects the widest companions. In the case of multi-companions systems, perturbations of outer companions can induce destabilization as a perturbation of outer component can propagate to the inner system \citep[e.g.,][]{hao2013, cai2015, cai2016}. The Galactic field itself also affects the evolution of wide binaries \citep[e.g.,][]{jiang2010, kaib2013}. Finally, other stars and brown dwarfs in the environment may be captured by the host system following a close encounter, and previously escaped secondaries may be captured by other stars as well \citep[e.g.,][]{kouwenhoven2010, perets2012}, which may provide an alternative solution to the origin of wide, low-mass companions. The stars and brown dwarfs in the solar neighbourhood likely represent a mixed population resulting from different formation mechanisms and from a different environmental interaction history. Despite the many unknowns, our model make clear predictions that can be statistically compared to nearby stars and brown dwarfs, to further constrain their origin.

\acknowledgments We wish to thank the anonymous referee for her/his very useful suggestions that helped to improve this paper. We are grateful to the University of Central Lancashire for support provided through the Distinguished Visitors Programme. Part of the activities relating to this work were supported by a China-Cardiff Centre Competition Fund, from Cardiff University.
M.B.N.K. was supported by the Peter and Patricia Gruber Foundation through the PPGF fellowship, by the Peking University One Hundred Talent Fund (985), and by the National Natural Science Foundation of China (grants 11010237, 11050110414, 11173004 and 11573004). This publication was made possible through the support of a grant from the John Templeton Foundation and National Astronomical Observatories of Chinese Academy of Sciences. The opinions expressed in this publication are those of the author(s) do not necessarily reflect the views of the John Templeton Foundation or National Astronomical Observatories of Chinese Academy of Sciences. The funds from John Templeton Foundation were awarded in a grant to The University of Chicago which also managed the program in conjunction with National Astronomical Observatories, Chinese Academy of Sciences. D.S. acknowledges support from STFC grant ST/M000877/1.

\end{document}